\documentclass[prl,twocolumn,floatfix,a4paper,superscriptaddress,round,comma]{revtex4-1}

\usepackage{bm,color,amsmath,txfonts}

\usepackage{graphicx}
\usepackage{subfigure}
\usepackage{verbatim}
\usepackage{dcolumn}
\usepackage{bm}
\usepackage{epsf}
\usepackage{xcolor}
\usepackage{hyperref}
\usepackage{hhline}
\usepackage{float}
\usepackage{enumerate}
\usepackage{bbm}
\usepackage{lipsum}
\usepackage{natbib}
\usepackage{soul}
\definecolor{mygreen}{rgb}{0.01, 0.31, 0.59}
\definecolor{myblue}{rgb}{0.01, 0.31, 0.59}
\definecolor{myred}{rgb}{0.63, 0.12, 0.12}
\hypersetup{
 colorlinks=true,   
 linkcolor=blue,  
 citecolor=blue, 
 urlcolor =blue    
}

\hypersetup{
 colorlinks=true,   
 linkcolor=blue,  
 citecolor=blue, 
 urlcolor =blue    
}

\renewcommand{\thefigure}{\arabic{figure}}

\begin{document}

\title{Mechanical Bistability in Kerr-modified Cavity Magnomechanics}

\author{Rui-Chang Shen}
\affiliation{Interdisciplinary Center of Quantum Information, State Key Laboratory of Modern Optical Instrumentation, and Zhejiang Province Key Laboratory of Quantum Technology and Device, School of Physics, Zhejiang University, Hangzhou 310027, China}
\author{Jie Li}\email{jieli007@zju.edu.cn}
\affiliation{Interdisciplinary Center of Quantum Information, State Key Laboratory of Modern Optical Instrumentation, and Zhejiang Province Key Laboratory of Quantum Technology and Device, School of Physics, Zhejiang University, Hangzhou 310027, China}
\author{Zhi-Yuan Fan}
\affiliation{Interdisciplinary Center of Quantum Information, State Key Laboratory of Modern Optical Instrumentation, and Zhejiang Province Key Laboratory of Quantum Technology and Device, School of Physics, Zhejiang University, Hangzhou 310027, China}
\author{Yi-Pu Wang}\email{yipuwang@zju.edu.cn}
\affiliation{Interdisciplinary Center of Quantum Information, State Key Laboratory of Modern Optical Instrumentation, and Zhejiang Province Key Laboratory of Quantum Technology and Device, School of Physics, Zhejiang University, Hangzhou 310027, China}
\author{J. Q. You}\email{jqyou@zju.edu.cn}
\affiliation{Interdisciplinary Center of Quantum Information, State Key Laboratory of Modern Optical Instrumentation, and Zhejiang Province Key Laboratory of Quantum Technology and Device, School of Physics, Zhejiang University, Hangzhou 310027, China}

\begin{abstract}
Bistable mechanical vibration is observed in a cavity magnomechanical system, which consists of a microwave cavity mode, a magnon mode, and a mechanical vibration mode of a ferrimagnetic yttrium-iron-garnet (YIG) sphere. The bistability manifests itself in both the mechanical frequency and linewidth under a strong microwave drive field, which simultaneously activates three different kinds of nonlinearities, namely, magnetostriction, magnon self-Kerr, and magnon-phonon cross-Kerr nonlinearities. The magnon-phonon cross-Kerr nonlinearity is first predicted and measured in magnomechanics. The system enters a regime where Kerr-type nonlinearities strongly modify the conventional cavity magnomechanics that possesses only a {radiation-pressure-like} magnomechanical coupling. Three different kinds of nonlinearities are identified and distinguished in the experiment. Our work demonstrates a new mechanism for achieving mechanical bistability by combining magnetostriction and Kerr-type nonlinearities, and indicates that such Kerr-modified cavity magnomechanics provides a unique platform for studying many distinct nonlinearities in a single experiment.
\end{abstract}
\maketitle

\vspace{.3cm}
\textit{Introduction.---}Bistability, or multistability, discontinuous jumps, and hysteresis are characteristic features of nonlinear systems. Bistability is a widespread phenomenon that exists in a variety of physical systems, e.g., optics~\cite{Walls1,Walls2,Lugiato}, electronic tunneling structures~\cite{Goldman}, magnetic nanorings~\cite{Zhu}, thermal radiation~\cite{Biehs}, a driven-dissipative superfluid~\cite{Ott}, and cavity magnonics~\cite{YP18}. Its presence requires nonlinearity in the system. To date, bistability has been studied in various mechanical systems, including nano- or micromechanical resonators~\cite{Capasso,Zant03,Badzey}, piezoelectric beams~\cite{Cottone}, mechanical morphing structures~\cite{Chen}, and levitated nanoparticles~\cite{Ricci}. Bistable mechanical motion finds many important applications: It is the basis for mechanical switches~\cite{switch1,switch2}, memory elements~\cite{Badzey2,Zant09}, logic gates~\cite{Murali}, vibration energy harvesters~\cite{Cottone,Harne}, and signal amplifiers~\cite{Badzey,Ricci}, etc.

Different mechanisms can bring about nonlinearity in the system leading to bistable mechanical motion. Most commonly, a strong drive can induce bistability of a mechanical oscillator, of which the dynamics is described by the Duffing equation~\cite{Badzey,Cleland,Buks,Katz}. {Mechanical bistability} can also be caused by the Casimir force~\cite{Capasso}, nanomechanical effects on Coulomb blockade~\cite{Zant03}, magnetic repulsion~\cite{Neri}, and intrinsic nonlinearity in the optomechanical coupling~\cite{Seok}, etc.

{Here} we introduce a mechanism to induce mechanical bistability, distinguished from all the above mechanisms, by exploiting rich nonlinearities in the ferrimagnetic yttrium-iron-garnet (YIG) in cavity magnomechanics (CMM).
In the CMM~\cite{Tang,APE,Davis,Yuan}, magnons are the quanta of collective spin excitations {in magnetically ordered materials, such as YIG. They} can strongly couple to microwave cavity photons by the magnetic-dipole interaction, leading to cavity polaritons~\cite{S1,S2,S3,S4,S5,S6}. They can also couple to deformation vibration phonons of the ferrimagnet via the magnetostrictive force~\cite{Kittle,Tang,Davis}. Such a {radiation-pressure-like} magnomechanical coupling provides necessary nonlinearity, enabling a number of theoretical proposals, including the preparation of entangled states~\cite{Jie18,Jie19b,Tan,Jie20,QST}, squeezed states~\cite{Jie19a,HFW,nsr}, mechanical quantum-ground states~\cite{Ding,HFW2,FRe}, slow light~\cite{Xiong,Ullah}, {thermometry}~\cite{Davis2}, quantum memory~\cite{JJ,Twamley}, exceptional points~\cite{JH}, and parity-time-related phenomena~\cite{YXL,Sun,Wu,Ding3}, etc. In contrast, the experimental studies on this system are, by now, very limited: Magnomechanically induced transparency and absorption~\cite{Tang} and mechanical cooling and lasing~\cite{Davis} {have been} demonstrated.

{In this Letter}, we report an experimental observation of bistable mechanical vibration of a YIG sphere in the CMM. We show that both the frequency and linewidth of the mechanical mode exhibit a bistable feature as a result of the combined effects of the {radiation-pressure-like} magnetostrictive interaction~\cite{Tang,Davis}, the magnon self-Kerr~\cite{YP16,YP21}, and the magnon-phonon cross-Kerr nonlinearities. Three different kinds of nonlinearities are simultaneously activated by applying a strong drive field on the YIG sphere.  Their respective contributions to the mechanical frequency and linewidth are discussed. 

\begin{figure}
 \centering
 \hskip-0.4cm\includegraphics{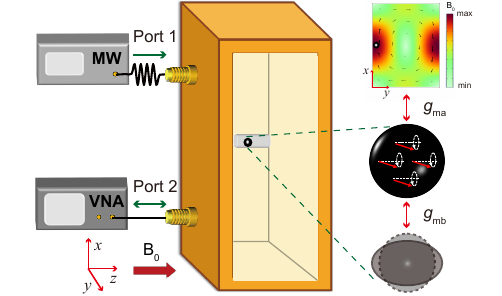}
 \caption{Device schematic. Left panel: Schematic of the CMM system. A 0.28 mm-diameter YIG sphere is placed (free to move) in a horizontal 0.9 mm-inner-diameter glass capillary and at the antinode of the magnetic field of the cavity mode $\rm{TE}_{\rm{101}}$. The cavity has two ports: Port 1 is connected to a microwave source (MW) to load the drive field, and Port 2 is connected to a vector network analyzer (VNA) to measure the reflection of the probe field with the power of $-5$ dBm. We set the direction of the bias magnetic field $B_{\rm{0}}$ as the $z$ direction, and the vertical direction as the $x$ direction. Right panel: Schematic of the coupled three modes.}
  \label{fig1}
\end{figure}

\textit{Kerr-modified CMM.---}The CMM system consists of a microwave cavity mode, a magnon mode, and a mechanical vibration mode, see Fig.~\ref{fig1}. {In the experiment, we use the oxygen-free copper cavity with dimensions of $42\times22\times8~{\rm{mm}^{3}}$. The cavity $\rm{TE}_{\rm{101}}$ mode has a frequency $\omega_{\rm{a}}/2\pi \,\,{=}\,\, 7.653~\rm{GHz}$, and a total decay rate $\kappa_{\rm{a}}/2\pi=2.78~\rm{MHz}$. The cavity decay rates associated with the two ports are $\kappa_{\rm{1,2}}/2\pi \,\,{=}\,\,0.22~\rm{MHz}$ and $1.05~\rm{MHz}$, respectively. The magnon and mechanical modes are supported by a 0.28 mm-diameter YIG sphere. The frequency of the magnon mode can be tuned by adjusting the bias magnetic field $B_{0}$ via $\omega_{\rm{m}}=\gamma B_{0}$, with $\gamma$ being the gyromagnetic ratio. The magnon dissipation rate is $\kappa_{\rm{m}}/2\pi=2.2~\rm{MHz}$. The magnon mode couples to the cavity magnetic field by the magnetic-dipole interaction with the coupling strength $g_{\rm{ma}}/2\pi=7.37~\rm{MHz}$, and to a vibration mode by the magnetostrictive (radiation-pressure-like) interaction with the bare magnomechanical coupling strength $g_{\rm{mb}}/2\pi \,\,{=}\,\, 1.22~\rm{mHz}$. Here we consider the lower-frequency mechanical mode (with a natural frequency $\omega_{\rm{b}}/2\pi \,\,{=}\,\,11.0308~\rm{MHz}$ and linewidth $\kappa_{\rm{b}}/2\pi \,\,{=}\,\, 550~\rm{Hz}$) in our observed two adjacent mechanical modes, which has a stronger coupling $g_{\rm{mb}}$. The magnomechanical coupling can be significantly enhanced by applying a pump field on the magnon mode~\cite{Jie18}. In our experiment, this is realized by strongly driving the cavity, which linearly couples to the magnon mode. In Ref.~\cite{SM}, we provide a list of parameters and the details of how they are extracted by fitting the experimental data.   }

Under a strong pump, the Hamiltonian of the CMM system is given by~\cite{SM}
\begin{equation}\label{f-1}
\begin{aligned}
H/\hbar &=\omega_{\rm{a}} a^{\dagger} a+\omega_{\rm{m}} m^{\dagger} m +\omega_{\rm{b}} b^{\dagger} b+g_{\rm ma}(a^{\dagger} m+a m^{\dagger})\\
&+g_{\rm mb} m^\dagger m  \left (b+ b^\dagger\right)+H_{\rm{Kerr}}/\hbar  +\!\! \sqrt{\kappa_{\rm{1}}}\varepsilon_{\rm{d}}\left(a^{\dagger} e^{-i \omega_{\rm{d}} t} + {\rm H.c.} \right),
\end{aligned}
\end{equation}
{where $a$, $m$, and $b$ ($a^{\dag}$, $m^{\dag}$, and $b^{\dag}$) are the annihilation (creation) operators of the cavity mode, the magnon mode, and the mechanical mode, respectively. }
The last term is the driving Hamiltonian, where $\kappa_1$ is the cavity decay rate associated with the driving port (Port 1), and $\varepsilon_{\rm{d}}=\sqrt{P_{\rm{d}}/(\hbar\omega_{\rm{d}})}$, with $P_{\rm{d}}$ ($\omega_{\rm{d}}$) being the power (frequency) of the microwave drive field.   The novel part of the Hamiltonian, with respect to the conventional CMM, is the Kerr nonlinear term $H_{\rm{Kerr}}$ activated by the strong pump field~\cite{SM}
\begin{eqnarray}\label{f-2}
H_{\rm{Kerr}}/\hbar&=&K_{\rm{m}} m^{\dagger} m m^{\dagger} m +K_{\rm{cross}}m^{\dagger} m  b^{\dagger} b,
\end{eqnarray}
where $K_{\rm{m}}$ is the magnon self-Kerr coefficient, and $K_{\rm{cross}}$ is the magnon-phonon cross-Kerr coefficient. The magnon self-Kerr effect is caused by the magnetocrystalline anisotropy~\cite{YP16,YP21}, and the cross-Kerr nonlinearity originates from the magnetoelastic coupling by including the second-order terms in the strain tensor~\cite{Landau,Tero},
\begin{equation}\label{f-3}
\epsilon_{ij}=\frac{1}{2}\left ( \frac{\partial u_i}{\partial l_j} +\frac{\partial u_j}{\partial l_i} +\sum_{k}\frac{\partial u_k}{\partial l_i}\frac{\partial u_k}{\partial l_j} \right ),
\end{equation}
where $u_i$ are the components of the displacement vector, {and $l_i = i$ ($i=x,y,z$)}. The first-order terms lead to the conventional {radiation-pressure-like} interaction Hamiltonian~\cite{Fan}, $\hbar g_{\rm mb} m^\dagger m  \left (b+ b^\dagger\right)$. Under a moderate drive field, the second-order terms are negligible~\cite{Tang,Davis}, but can no longer be neglected when the drive becomes sufficiently strong, as in our experiment, yielding an appreciable magnon-phonon cross-Kerr nonlinearity. As will be seen later, the cross-Kerr nonlinearity is indispensable in the model for fitting the mechanical frequency shift.

Both the magnon self-Kerr and magnon-phonon cross-Kerr terms, as well as the {radiation-pressure-like} term, cause a magnon frequency shift $\delta\omega_{\rm{m}}=2 K_{\rm{m}}|M|^{2} + K_{\rm{cross}}|B|^{2} + 2 g_{\rm mb} {\rm Re}[B]$, where $O\,{=}\,\langle o \rangle$ ($o\,{=}\, m, b, a$) denote the average of the modes. In our experiment, the dominant contribution is from the self-Kerr nonlinearity~\cite{SM}, which gives a bistable magnon frequency shift~\cite{YP18}. Also, the cross-Kerr nonlinearity causes a mechanical frequency shift $ \delta\omega_{\rm{b}} = K_{\rm{cross}}|M|^{2} $. Using the Heisenberg-Langevin approach, we obtain the equation for the steady-state average $M$~\cite{SM}
\begin{eqnarray}\label{f-4}
\begin{aligned}
\eta_{\rm{a}} \kappa_{\rm{1}} & g_{\rm{ma}}^2  \varepsilon_{\rm{d}}^2 =  |M|^{2}  \times  \\
&\left[\left(\Delta_{\rm{m}}-\eta_{\rm{a}}g_{\rm{ma}}^2 \Delta_{\rm{a}} +2 K_{\rm{m}}|M|^{2} \right)^2+ \left(\frac{\kappa_{\rm{m}}}{2}+\eta_{\rm{a}}g_{\rm{ma}}^2 \frac{\kappa_{\rm{a}}} {2} \right)^2 \right],
\end{aligned}
\end{eqnarray}
where $\Delta_{\rm{a\,(m)}} =\omega_{\rm{a\,(m)}}- \omega_{\rm{d}}$, $\eta_{\rm{a}}= \frac{1} {\Delta_{\rm{a}}^2+(\kappa_{\rm{a}}/2)^2}$. 
In deriving Eq.~(\ref{f-4}), we neglect contributions from the mechanical mode to the magnon frequency shift, because of a much smaller mechanical excitation number compared to the magnon excitation number for the drive powers used in this work. It is a cubic equation of the magnon excitation number $|M|^2$. In a suitable range of the drive power, there are two stable solutions, leading to the bistable magnon and phonon frequency shifts by varying the drive power.

\begin{figure}[t]
 \centering
 \hskip-0.4cm\includegraphics{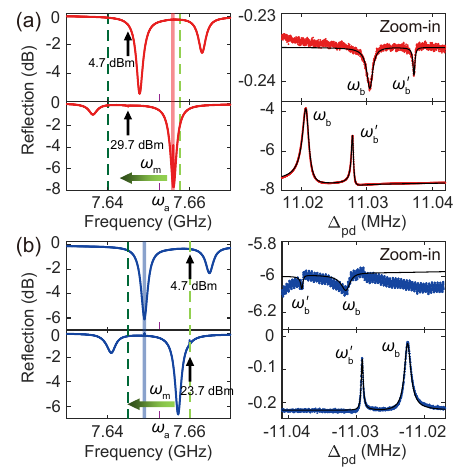}
 \caption{(a) Left panel: Measured reflection spectra under a red-detuned drive. The frequency of the drive field $\omega_{\rm{d}}/2\pi=7.645~\rm{GHz}$ (black arrow). By increasing the drive power, the magnon frequency shift is negative: $\omega_{\rm{m}}^{\rm{L}}/2\pi=7.658~\rm{GHz}$ {(light green dashed line) at a lower power} $P_{\rm{d}}=4.7~\rm{dBm}$ (upper panel) and $\omega_{\rm{m}}^{\rm{H}}/2\pi=7.640~\rm{GHz}$ {(dark green dashed line)} at $P_{\rm{d}}=29.7~\rm{dBm}$ (lower panel). The green arrow indicates the direction in which the magnon frequency shifts by increasing the power.  Right panel: Zoom-in on the red shaded areas in the left panel shows detailed spectra of the magnomechanically induced resonances, where $\Delta_{\rm{pd}}=\omega_{\rm{p}}-\omega_{\rm{d}}$. The black lines are the fitting curves. (b) Left panel: Measured reflection spectra under a blue-detuned drive. {The drive frequency $\omega_{\rm{d}}/2\pi=7.660~\rm{GHz}$. By adjusting the bias magnetic field, the magnon frequency is tuned close to the drive frequency $\omega_{\rm{m}}^{\rm{L}}/2\pi \simeq \omega_{\rm{d}}$ at the power $P_{\rm{d}}=4.7~\rm{dBm}$}. Increasing the power to $23.7~\rm{dBm}$, the magnon frequency $\omega_{\rm{m}}^{\rm{H}}/2\pi=7.645~\rm{GHz}$. Right panel: Zoom-in on the blue shaded areas in the left panel shows detailed spectra of the magnomechanically induced resonances. {We observe two adjacent mechanical modes with the frequencies $\omega_{\rm{b}}/2\pi \,\,{=}\,\,11.0308~\rm{MHz}$ and $\omega'_{\rm{b}}/2\pi=11.0377~\rm{MHz}$.} Due to their similar behaviors, we focus on the lower-frequency mode in the text. } 
  \label{fig2}
\end{figure}

The {radiation-pressure-like} coupling gives rise to an effective susceptibility of the mechanical mode~\cite{SM}
\begin{eqnarray}\label{f-5}
\begin{aligned}
\chi_{\rm{b,eff}}(\omega)
=\left( \chi_{\rm{b}}^{-1}(\omega)-2i|G_{\rm{mb}}|^{2}\left( \chi_{\rm{ma}}(\omega)- \chi_{\rm{ma}}^{*}(-\omega) \right) \right)^{-1},
\end{aligned}
\end{eqnarray}
where $\chi_{\rm{b}}(\omega)$ is the natural susceptibility of the mechanical mode, but depends on the modified mechanical frequency $\tilde{\omega}_{\rm{b}}=\omega_{\rm{b}}+ K_{\rm{cross}}|M|^{2} $, which includes the cross-Kerr induced frequency shift. The effective coupling $G_{\rm{mb}}=g_{\rm{mb}}M$, and $\chi_{\rm{ma}}(\omega)=\left[ \chi_{\rm{m}}^{-1}(\omega)+g_{\rm{ma}}^2 \chi_{\rm{a}}(\omega) \right]^{-1}$, where $\chi_{\rm{m}}(\omega)$ and $\chi_{\rm{a}}(\omega)$ are the natural susceptibilities of the magnon and cavity modes, with the magnon detuning in $\chi_{\rm{m}}(\omega)$ modified as $\tilde{\Delta}_{\rm{m}}=\Delta_{\rm{m}}+2 K_{\rm{m}}|M|^{2}$, which includes the dominant magnon self-Kerr induced frequency shift. See \cite{SM} for the explicit expressions of the susceptibilities.

The effective mechanical susceptibility yields a frequency shift of the phonon mode (the so-called ``magnonic spring" effect~\cite{Davis}, in analogy to the ``optical spring" in optomechanics~\cite{MA})
\begin{equation}\label{f-6}
\begin{aligned}
\delta\omega_{\rm{b}}= -{\rm{Re}}\left[ 2i|G_{\rm{mb}}|^{2}\left( \chi_{\rm{ma}}(\omega)- \chi_{\rm{ma}}^{*}(-\omega)\right) \right]+ K_{\rm{cross}} |M|^{2},
\end{aligned}
\end{equation}
where we write together the frequency shift induced by the cross-Kerr nonlinearity. Moreover, it leads to a mechanical linewidth change
\begin{equation}\label{f-7}
\begin{aligned}
\delta\Gamma_{\rm{b}}={\rm{Im}}\left[ 2i|G_{\rm{mb}}|^{2}\left( \chi_{\rm{ma}}(\omega)- \chi_{\rm{ma}}^{*}(-\omega)\right) \right].
\end{aligned}
\end{equation}
Clearly, this {\it linewidth} change is only caused by the {radiation-pressure-like} coupling, distinguished from the {\it frequency} shift caused by the self-Kerr or cross-Kerr nonlinearity. By applying a red- or blue-detuned drive field, we can choose to operate the system in two different regimes, where either the magnomechanical anti-Stokes or Stokes scattering is dominant. This yields an increased ($\delta\Gamma_{\rm{b}}>0$) or a reduced ($\delta\Gamma_{\rm{b}}<0$) mechanical linewidth, corresponding to the cooling or amplification of the mechanical motion~\cite{Davis,MA}.

In our system, due to the strong coupling $g_{\rm{ma}} > \kappa_{\rm{m}}, \kappa_{\rm{a}}$, the magnon and cavity modes form two cavity polariton (hybridized) modes (Fig.~\ref{fig2}, left panels)~\cite{S1,S2,S3,S4,S5,S6}. Here, a red (blue)-detuned drive means that the drive frequency is lower (higher) than the frequency of the cavity-like polariton mode, i.e., the ``deeper" polariton in the spectra close to the cavity resonance. For the red (blue)-detuned drive, we show the anti-Stokes (Stokes) sidebands associated with two mechanical modes for two drive powers in the zoom-in plots of Fig.~\ref{fig2}(a) (Fig.~\ref{fig2}(b)).  When the detuning between the drive field and the deeper polariton matches the mechanical frequencies, the anti-Stokes (Stokes) sidebands are manifested as the magnomechanically induced transparency (absorption)~\cite{Tang}.

\begin{figure}[b]
	\hskip-0.19cm\includegraphics[width=0.9\linewidth]{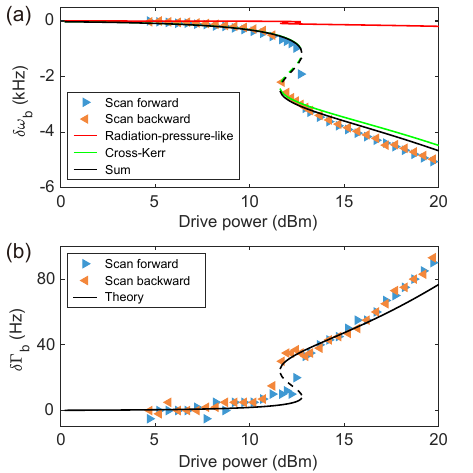}
	\caption{Bistable mechanical frequency and linewidth under a red-detuned drive. (a) The mechanical frequency shift versus the drive power. The red (green) curve is the fitting of the frequency shift induced by the {radiation-pressure-like} coupling (cross-Kerr effect) using Eq.~\eqref{f-6}, and the black curve is the sum of the two contributions. (b) The mechanical linewidth variation versus the drive power. The black curve is the fitting of the linewidth change using Eq.~\eqref{f-7}. In both figures, the blue (orange) triangles are the experimental data obtained via forward (backward) sweep of the drive power. }
	\label{fig3}
\end{figure}


{\textit{Red-detuned drive.---}To implement a red-detuned drive,} we drive the cavity with a microwave field at frequency $\omega_{\rm{d}}/2\pi=7.645~\rm{GHz}$. By adjusting the bias magnetic field, we tune the magnon frequency to be $\omega^{\rm L}_{\rm{m}}/2\pi \,\,{=}\,\, 7.658~\rm{GHz}$ at the drive power $P_{\rm{d}} \,\,{=}\,\, 4.7~\rm{dBm}$ (see Fig.~\ref{fig2}(a)). We have the [110] axis of the YIG sphere aligned parallel to the static magnetic field, which yields a negative self-Kerr coefficient $K_{\rm{m}}/2\pi \,\,{=}\,{-}6.5~\rm{nHz}$. {An increase in power thus results in a negative magnon frequency shift $\delta\omega_{\rm{m}}=2 K_{\rm{m}}|M|^{2}$,} and the magnon frequency reduces to $\omega^{\rm H}_{\rm{m}}/2\pi \,\,{=}\,\, 7.640 ~\rm{GHz}$ when the power increases to $P_{\rm{d}} \,\,{=}\,\, 29.7~\rm{dBm}$ (Fig.~\ref{fig2}(a)), {which yields an effective coupling $G_{\rm{mb}}/2\pi=45.8~\rm{kHz}$}. Under these conditions, the magnon excitation number $|M|^2$ shows a bistable behavior {through variation of the power.}

Equation~\eqref{f-6} indicates that the {radiation-pressure-like} coupling results in a mechanical frequency shift, and so does the cross-Kerr nonlinearity. This is confirmed by the experimental data in Fig.~\ref{fig3}(a). It shows that the cross-Kerr plays a dominant role because of a large magnon excitation number, and both the frequency shifts caused, respectively, by the cross-Kerr and the {radiation-pressure-like} coupling show a bistable feature with the forward and backward sweeps of the drive power. This is because both of them originate from the bistable magnon excitation number $|M|^2$, c.f. Eq.~\eqref{f-6}. Note that for the spring effect, the bistability of $|M|^2$ is mapped to the magnon frequency shift $\tilde{\Delta}_{\rm{m}}$, then to the polariton susceptibility $\chi_{\rm{ma}}(\omega)$, and finally to the mechanical frequency.

By increasing the drive power from $4.7~\rm{dBm}$ to $19.7~\rm{dBm}$, the cross-Kerr causes a maximum frequency shift of $-4.6~\rm{kHz}$ (Fig.~\ref{fig3}(a), the green line). The fitting cross-Kerr coefficient is $K_{\rm{cross}}/2\pi=-5.4~\rm{pHz}$. Under the red-detuned drive, the magnonic spring effect yields a negative frequency shift $\delta\omega_{\rm{b}} = \rm{Re}\left[ \chi_{\rm{b,eff}}^{-1}(\omega) -\chi_{\rm{b}}^{-1}(\omega)\right] <0$ (Fig.~\ref{fig3}(a), the red line), and the maximum frequency shift is $-200~\rm{Hz}$. Adding up these two frequency shifts gives the total mechanical frequency shift (Fig.~\ref{fig3}(a), the black line), which fits well with the experimental data (Fig.~\ref{fig3}(a), triangles) when the power is not too strong.

Another interesting finding is the bistable feature of the mechanical linewidth (Fig.~\ref{fig3}(b)). The magnomechanical backaction leads to the variation of the mechanical linewidth $\delta\Gamma_{\rm{b}}=-\rm{Im}\left[ \chi_{\rm{b,eff}}^{-1}(\omega) -\chi_{\rm{b}}^{-1}(\omega)\right]$. For a red-detuned drive, the anti-Stokes process is dominant, resulting in an increased mechanical linewidth $\delta\Gamma_{\rm{b}} >0$ and the cooling of the motion. The bistable mechanical linewidth is also induced by the bistable $|M|^2$ (see Eq.~\eqref{f-7}), similar to the mechanical frequency. The theory fits well with the experimental results, and the discrepancy appears only in the high-power regime. {This is because the considerable heating effect at strong pump powers can broaden the mechanical linewidth~\cite{heat1,heat2}, which is not included in our model. }

\textit{Blue-detuned drive.---}When a blue-detuned drive is applied, the system enters a regime where the Stokes scattering is dominant. The magnomechanical parametric down-conversion amplifies the mechanical motion with the characteristic of a reduced linewidth. Furthermore, the mechanical frequency shift induced by the spring effect will move in the opposite direction compared with the red-detuned drive.

To implement a blue-detuned drive, we drive the cavity with a microwave field at frequency $\omega_{\rm{d}}/2\pi=7.66~\rm{GHz}$, and tune the magnon frequency close to the drive frequency (see Fig.~\ref{fig2}(b)). We attempted to make the Stokes sideband of the drive field resonate with the ``deeper" polariton {\it at a high pump power}, such that the Stokes scattering rate is maximized and the magnomechanical coupling strength $G_{\rm mb}$ becomes strong.  However, to meet the drive conditions for a {\it bistable} magnon excitation number $|M|^2$, the drive frequency is restricted to a certain range, which hinders us to make the Stokes sideband and the ``deeper" polariton resonate. Therefore, we only achieve this at lower drive powers, giving a faint magnomechanically induced absorption (Fig.~\ref{fig2}(b), upper panels).

From the red to the blue detuning, we only adjust the magnon and the drive frequencies. Because the direction of the crystal axis is unchanged, the magnon self-Kerr coefficient $K_{\rm{m}}$ is still negative, so again a negative frequency shift by increasing the power (green arrow in Fig.~\ref{fig2}(b)). For the power up to 23.7 dBm, which yields $G_{\rm{mb}}/2\pi=42.7~\rm{kHz}$, the frequency of the cavity-like polariton is always lower than the drive frequency, so the system is operated under a blue-detuned drive.

\begin{figure}[t]
	\hskip-0.19cm\includegraphics[width=0.9\linewidth]{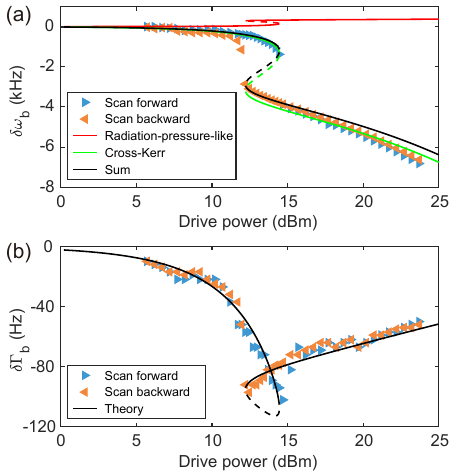}
\caption{Bistable mechanical frequency and linewidth under a blue-detuned drive. (a) The mechanical frequency shift and (b) the mechanical linewidth change versus the drive power. The curves and the triangles are shown in the same manner as in Fig.~\ref{fig3}. }
\label{fig4}
\end{figure}

Figure~\ref{fig4} displays the bistable mechanical frequency shift $\delta\omega_{\rm{b}}$ and linewidth change $\delta\Gamma_{\rm{b}}$. For the frequency shift, both the contributions from the cross-Kerr and the {radiation-pressure-like} coupling should be considered, but the former plays a dominant role (Fig.~\ref{fig4}(a), the green line), as in the case of the red-detuned drive, yielding a frequency shift of $-6.5$ kHz at the power of 23.7 dBm. Differently, the spring effect induced frequency shift (370 Hz at $23.7$ dBm) is positive (Fig.~\ref{fig4}(a), the red line). The opposite frequency shifts by the spring effect in the blue and red-detuned drives agree with the finding of Ref.~\cite{Davis}, but no bistability was observed in their work.

The reduced mechanical linewidth $\delta\Gamma_{\rm{b}}<0$ under a blue-detuned drive is confirmed by Fig.~\ref{fig4}(b). However, unlike the bistable curve in the red-detuned drive case (Fig.~\ref{fig3}(b)), $\delta\Gamma_{\rm{b}}$ manifests the bistability in an {{\it alpha}-shaped curve} by sweeping the drive power.  This is the result of the trade-off between the growing coupling strength $G_{\rm mb}$ (which enhances the Stokes scattering rate, yielding an increasing $|\delta\Gamma_{\rm{b}}|$) and the larger detuning between the Stokes sideband and the ``deeper" polariton (Fig.~\ref{fig2}(b)) (which reduces the Stokes scattering rate, resulting in a decreasing $|\delta \Gamma_{\rm{b}}|$) by raising the drive power. These two effects are balanced when the power is in the range of 12 dBm to 15 dBm.

\textit{Conclusions.---}We have observed bistable mechanical frequency and linewidth and the magnon-phonon cross-Kerr nonlinearity in the CMM system. The mechanical bistability results from the magnomechanical backaction on the mechanical mode and the strong modifications on the backaction due to the magnon self-Kerr and magnon-phonon cross-Kerr nonlinearities. {The effects of the magnon self-Kerr, the magnon-phonon cross-Kerr, and the radiation-pressure-like interactions can be identified by measuring primarily} the magnon frequency shift, the mechanical frequency shift, and the mechanical linewidth, respectively. {The new mechanism for achieving bistable mechanical motion revealed by this work} promises a wide range of applications, such as in mechanical switches, memories, logic gates, and signal amplifiers.


\textit{Acknowledgments.} This work was supported by the National Natural Science Foundation of China (Grants Nos. 11934010, U1801661, 12174329, 11874249), Zhejiang Province Program for Science and Technology (Grant No. 2020C01019), and the Fundamental Research Funds for the Central Universities (No. 2021FZZX001-02).

\setcounter{figure}{0}
\setcounter{equation}{0}
\setcounter{table}{0}
\renewcommand\theequation{S\arabic{equation}}
\renewcommand\thefigure{S\arabic{figure}}
\renewcommand\thetable{S\arabic{table}}

\setcounter{section}{0}

\clearpage    
\onecolumngrid  

\section*{Supplemental Materials } 

\section{\uppercase\expandafter{\romannumeral1}.~SYSTEM PARAMETERS}

	\begin{table}[htbp]
	\centering
	\begin{tabular}{| c | c | c | }
		\hline
		Quantity & Symbol & Value\\
		\hline
        \hline
		 Gyromagnetic ratio & $\gamma$ & $2\pi \times$2.8 MHz/Oe \\
		\hline
		 Frequency of the cavity $\rm{TE}_{\rm{101}}$ mode & $\omega_{\rm{a}}$ & $2\pi \times$7.653 GHz \\
		\hline
                 Frequency of the magnon mode & $\omega_{\rm{m}}$ & $\gamma B_0$ \\
		\hline
		 Frequency of the lower-frequency phonon mode & $\omega_{\rm{b}}$ & $2\pi \times$11.0308 MHz \\
                 \hline
		 Frequency of the higher-frequency phonon mode & $\omega'_{\rm{b}}$ & $2\pi \times$11.0377 MHz \\
		\hline
                Total cavity decay rate & $\kappa_{\rm{a}}$ & $2\pi \times$2.78 MHz \\
              \hline
              Cavity decay rate via Port 1 & $\kappa_{\rm{1}}$ & $2\pi \times$0.22 MHz \\
              \hline
              Cavity decay rate via Port 2  & $\kappa_{\rm{2}}$ & $2\pi \times$1.05 MHz \\
	      \hline
               Linewidth of the magnon mode & $\kappa_{\rm{m}}$ & $2\pi \times$2.2 MHz \\
	       \hline
	       Linewidth of the lower-frequency phonon mode & $\kappa_{\rm{b}}$ & $2\pi \times$550 Hz \\
               \hline
	       Linewidth of the higher-frequency phonon mode & $\kappa'_{\rm{b}}$ & $2\pi \times$180 Hz \\
                \hline
                Cavity-magnon coupling strength & $g_{\rm{ma}}$ & $2\pi \times$7.37 MHz \\
		\hline
		 Bare magnomechanical coupling strength (for the lower-frequency mode) & $g_{\rm{mb}}$ & $2\pi \times$1.22 mHz \\
                \hline
		 Bare magnomechanical coupling strength (for the higher-frequency mode) & $g'_{\rm{mb}}$ & $2\pi \times$0.62 mHz \\
               \hline
                Magnon self-Kerr coefficient  & $K_{\rm{m}}$ & -$2\pi \times$6.5 nHz \\
		\hline
		Magnon-phonon cross-Kerr coefficient & $K_{\rm{cross}}$ & -$2\pi \times$5.4 pHz \\
		\hline
	\end{tabular}
	\caption{List of system parameters.}
	\label{tab:1}
\end{table}

{Table~\ref{tab:1} provides a list of the system parameters. In the experiment, the frequencies and the dissipation rates (linewidths) of the cavity and magnon modes and their coupling strength $g_{\rm{ma}}$ are extracted by fitting the cavity-magnon polariton in the reflection spectra using Eq. \eqref{sup30}. The mechanical frequency and linewidth and the magnomechanical coupling strength $g_{\rm{mb}}$ are obtained by fitting the spectra of the magnomechanically induced resonances using also Eq. \eqref{sup30}. The cavity decay rates $\kappa_{\rm{1}}$ and $\kappa_{\rm{2}}$ associated with the two ports need to be fitted by measuring the reflection spectrum through the two ports, respectively. The magnon self-Kerr coefficient $K_{\rm{m}}$ of the crystal axis [110] is determined by measuring the magnon frequency shift, which is consistent with the value calculated from Eq. \eqref{sup4}.  The magnon-phonon cross-Kerr coefficient $K_{\rm{cross}}$ is obtained by fitting the mechanical frequency shift $\delta\omega_{\rm{b}}$ using Eq. \eqref{sup26}. Specifically, the first term of Eq. \eqref{sup26} is the frequency shift caused by the magnon-phonon radiation-pressure-like coupling, which can be calculated (e.g., -200 Hz under a red-detuned drive with the power 19.7 dBm). The magnon excitation number $|M|^{2}$ can also be calculated by using Eq. \eqref{sup20}. With all these at hand, the cross-Kerr coefficient $K_{\rm{cross}}$ can then be determined by fitting the mechanical frequency shift. }

\section{\uppercase\expandafter{\romannumeral2}.~Derivation of the Hamiltonian for Kerr nonlinearities}
In this section, we provide a detailed derivation of the Hamiltonian $H_{\rm Kerr}$ associated with the two Kerr nonlinear terms, namely, the magnon self-Kerr nonlinearity and the magnon-phonon cross-Kerr nonlinearity. These two terms become appreciable in the system when the pump field is sufficiently strong, such that the system enters a regime where the Kerr-type nonlinearities strongly modify the conventional cavity magnomechanics (CMM), which we term as the Kerr-modified CMM. Here the conventional CMM means that there is only a {radiation-pressure-like} coupling between magnons and vibration phonons~\cite{Tang,Davis}.

\subsection{A.~Magnon self-Kerr nonlinearity}

The magnon self-Kerr nonlinearity originates from the anisotropic field. When the bias magnetic field is aligned along the [110] axis of the YIG sphere, the anisotropic field is given by~\cite{Gurevich}
\begin{equation}\label{sup1}
\mathbf{H}_{\rm{an}}=\frac{3K_{\rm{an}}M_{x}}{\mu_{0}M^{2}}\mathbf{e}_{x}+ \frac{9K_{\rm{an}}M_{y}}{4\mu_{0}M^{2}}\mathbf{e}_{y}+ \frac{K_{\rm{an}}M_{z}}{\mu_{0}M^{2}}\mathbf{e}_{z},
\end{equation}
where $K_{\rm{an}}$ is the first-order magnetocrystalline anisotropy constant, and for the YIG at room temperature $K_{\rm{an}}=-610~\rm{J}/\rm{m}^3$. $\mathbf{M}=(M_{x},M_{y},M_{z})$ denotes the magnetization of the YIG sphere, $M$ is the saturation magnetization, and $\mu_{0}$ is the permeability of vacuum. The anisotropy Hamiltonian reads
\begin{equation}\label{sup2}
\begin{aligned}
H_{\rm{an}}&=-\frac{\mu_{0}}{2}\int_{V_{\rm{m}}}\mathbf{M}\cdot\mathbf{H}_{\rm{an}}d\tau ,\\
&=-\frac{K_{\rm{an}}V_{\rm{m}}}{8 M^{2}}\left(12M_{x}^{2}+9M_{y}^{2}+4M_{z}^{2}\right),
\end{aligned}
\end{equation}
where $V_{\rm{m}}$ is the volume of the YIG sphere. By using the relation $\mathbf{S}=\mathbf{M}V_{\rm{m}}/\hbar\gamma\equiv(S_{x},S_{y},S_{z})$~\cite{Soykal10}, with $\mathbf{S}$ being the macrospin operator and $\gamma$ being the gyromagnetic ratio, the anisotropy Hamiltonian $H_{\rm{an}}$ can be written as
\begin{equation}\label{sup3}
\begin{aligned}
H_{\rm{an}}/\hbar=-\frac{3 \hbar K_{\rm{an}}\gamma^2}{2M^{2}V_{\rm{m}}}S_{x}^2- \frac{9 \hbar K_{\rm{an}}\gamma^2}{8M^{2}V_{\rm{m}}}S_{y}^2- \frac{ \hbar K_{\rm{an}}\gamma^2}{2M^{2}V_{\rm{m}}}S_{z}^2.
\end{aligned}
\end{equation}

Using the Holstein-Primakoff transformation~\cite{Holstein40}: $S^{+}=\sqrt{2 S-m^{\dagger} m}m$, $S^{-}=m^{\dagger} \sqrt{2 S-m^{\dagger} m}$, and $S_{z}=S-m^{\dagger}m$, where $S$ is the total spin number of the macrospin and $m^{\dag}(m)$ is the creation (annihilation) operator of the magnon mode, we obtain
\begin{equation}\label{sup4}
\begin{aligned}
H_{\rm{an}}/\hbar\approx -\frac{13\hbar S K_{\rm{an}}\gamma^{2}}{8M^{2}V_{\rm{m}}} m^{\dagger} m + \frac{13\hbar K_{\rm{an}}\gamma^{2}}{16M^{2}V_{\rm{m}}} m^{\dagger} m m^{\dagger} m.
\end{aligned}
\end{equation}
The first term of the anisotropy Hamiltonian will modify the magnon frequency $\omega_{\rm{m}}^{'}=\omega_{\rm{m}}-\frac{13\hbar S K_{\rm{an}}\gamma^{2}}{8M^{2}V_{\rm{m}}}$, and the second term accounts for the magnon self-Kerr nonlinearity, which can be written in the form of $H_{\rm{self\textendash Kerr}}/\hbar = K_{\rm{m}} m^{\dagger} m m^{\dagger} m$, with the self-Kerr coefficient $K_{\rm{m}}=\frac{13\hbar K_{\rm{an}}\gamma^{2}}{16M^{2}V_{\rm{m}}}$. Note that in deriving the Hamiltonian \eqref{sup4}, we have omitted the constant term and the higher order terms.

\subsection{B.~Magnon-phonon cross-Kerr nonlinearity}

The magnetoelastic coupling describes the interaction between the magnetization and the elastic strain of the magnetic material. Depending on the distance between magnetic atoms (or ions), there are different kinds of interactions: the spin-orbital interaction, the exchange interaction between magnetic atoms (or ions), and the magnetic dipole-dipole interaction~\cite{Gurevich}.  In a cubic crystal, the magnetoelastic energy density is given by~\cite{Kittel49}
\begin{equation}\label{sup5}
\begin{split}
	f_{\rm{me}}=&\ \frac{b_1}{M^2}\left( M_x^2\epsilon_{xx} +M_y^2\epsilon_{yy}+M_z^2\epsilon_{zz}\right) +\frac{2b_2}{M^2}\left( M_xM_y\epsilon_{xy}+M_xM_z\epsilon_{xz}+M_yM_z\epsilon_{yz}\right),
\end{split}
\end{equation}
where $b_1$ and $b_2$ are the magnetoelastic coupling constants, and the strain tensor $\epsilon_{ij}$ is given in the nonlinear Euler-Bernoulli theory by~\cite{Landau,Tero}
\begin{equation}\label{sup6}
\epsilon_{ij}=\frac{1}{2}\left ( \frac{\partial u_i}{\partial l_j}+\frac{\partial u_j}{\partial l_i} +\sum_{k}\frac{\partial u_k}{\partial l_i}\frac{\partial u_k}{\partial l_j} \right ),
\end{equation}
with $u_i$ being the components of the displacement vector. The two first-order terms in $\epsilon_{ij}$ lead to the magnon-phonon {\it {radiation-pressure-like}} coupling (see \cite{Fan} for a strict derivation). In typical CMM experiments using a moderate drive field~\cite{Tang,Davis}, the second-order terms are neglected. However, for an intense drive field as used in our experiment, those terms will produce a noticeable effect. As we will derive below, those second-order terms in the strain tensor are responsible for the magnon-phonon cross-Kerr nonlinearity.

In the magnetoelastic energy density \eqref{sup5}, the second term accounts for the parametric magnon generation when the phonon frequency is twice the magnon frequency, or the linear magnon-phonon coupling when they are nearly resonant~\cite{Tang,Fan}. Thus, this term is negligible for our system with a low mechanical frequency $\omega_{\rm b} \ll \omega_{\rm m}$, and we can only consider the first term in \eqref{sup5}. Integrating over the whole volume of the YIG sphere, the interaction Hamiltonian can be written as
\begin{equation}\label{sup7}
\begin{aligned}
H_1 &= \frac{b_1}{M^2}\int dl^3\left (M_x^2 \epsilon_{xx}+ M_y^2\epsilon_{yy} +M_z^2\epsilon_{zz}\right ),\\
&\simeq \frac{b_1}{M}\frac{\hbar\gamma }{V_m} m^\dagger m\int dl^3\left ( \epsilon_{xx}+\epsilon_{yy}-2\epsilon_{zz}\right ).
\end{aligned}
\end{equation}

To quantize the above Hamiltonian, we express the displacement vector $u_i$ as a superposition
\begin{equation}\label{sup8}
	\vec{u}=\sum_{n,m,l} d^{\left( n,m,l\right) }\vec{\chi}^{(n,m,l)}\left(x,y,z \right),
\end{equation}
where $\vec{\chi}^{(n,m,l)}\left(x,y,z \right)$ is the displacement eigenmode with the corresponding amplitude $d^{\left( n,m,l\right)}$, and the superscript $ n,m,l$ denote the mode indices.  The mechanical displacement can be quantized as $d^{\left( n,m,l\right)}= d_{\rm{zpm}}^{\left( n,m,l\right)} (b_{n,m,l} + b^\dagger_{n,m,l})$, where $d_{\rm{zpm}}^{\left( n,m,l\right)}$ is the amplitude of the zero-point motion, and $b_{n,m,l}$ ($b^\dagger_{n,m,l}$) is the bosonic annihilation (creation) operator of the mechanical mode. Substituting \eqref{sup8} into the Hamiltonian \eqref{sup7}, we obtain
\begin{equation}\label{sup9}
\begin{aligned}
H_1\simeq\sum_{n,m,l}\hbar g_{\rm mb}^{\left ( n,m,l\right )} m^\dagger m  \left (b_{n,m,l}+ b^\dagger_{n,m,l}\right )+\sum_{n,m,l}\hbar K_{\rm cross}^{\left ( n,m,l\right )}  m^\dagger m b^\dagger_{n,m,l}b_{n,m,l},
\end{aligned}
\end{equation}
where we use the rotating-wave approximation by neglecting the fast-oscillating terms in deriving the second term. This is valid when $\omega_{\rm b} \gg K_{\rm cross} |M|^2 $, which is well satisfied in our experiment. The first term describes the magnon-phonon {radiation-pressure-like} interaction, and the second term accounts for the cross-Kerr interaction between the magnon and mechanical modes. The {radiation-pressure-like} coupling strength $g_{\rm mb}^{\left ( n,m,l\right )}$ and the cross-Kerr coefficient $K_{\rm cross}^{\left ( n,m,l\right )}$ are given by
\begin{equation}\label{sup10}
\begin{aligned}
g_{\rm mb}^{\left ( n,m,l\right )}=& \frac{b_1}{M} \frac{\gamma}{V_m}  \int dl^3  d_{\text{zpm}}^{\left( n,m,l\right) }\left ( \frac{\partial \chi_{x}^{(n,m,l)}}{\partial x}+\frac{\partial \chi_{y}^{(n,m,l)}}{\partial y}-2\frac{\partial \chi_{z}^{(n,m,l)}}{\partial z}\right ), \\
K_{\rm cross}^{\left ( n,m,l\right )}=&\frac{b_1}{M} \frac{\gamma}{V_m}  \int dl^3  d_{\text{zpm}}^{\left( n,m,l\right)^2 } \sum_{k}\left[ \left (\frac{\partial \chi_{k}^{(n,m,l)}}{\partial x}\right)^2 +\left (\frac{\partial \chi_{k}^{(n,m,l)}}{\partial y}\right)^2 -2\left (\frac{\partial \chi_{k}^{(n,m,l)}}{\partial z}\right)^2 \right].
\end{aligned}
\end{equation}

When considering a specific mechanical mode as in our experiment, the interaction Hamiltonian takes a simple form of
\begin{equation}\label{sup11}
\begin{aligned}
H_1/\hbar= g_{\rm mb} m^\dagger m  \left (b+ b^\dagger\right ) + K_{\rm cross}  m^\dagger m b^\dagger b.
\end{aligned}
\end{equation}

\section{\uppercase\expandafter{\romannumeral3}.~Hamiltonian of the Kerr-modified cavity magnomechanical system}

The CMM system under study consists of a microwave cavity mode, a magnon (Kittel) mode, and a mechanical vibration mode. The magnon mode couples to the cavity mode by the magnetic-dipole interaction, and to the mechanical mode by the magnetostrictive interaction. There is no direct coupling between the cavity and the mechanics. The drive field is applied to the cavity mode via the Port 1 of the cavity, and the probe field is sent via the Port 2 of the cavity. Under a strong pump, the magnon self-Kerr and magnon-phonon cross-Kerr nonlinearities are activated in the system. The total Hamiltonian of the CMM system is given by
\begin{equation}\label{sup12}
\begin{aligned}
H/\hbar \, =\, &\omega_{\rm{a}} a^{\dagger} a+\omega_{\rm{m}} m^{\dagger} m+\omega_{\rm{b}} b^{\dagger} b+g_{\rm ma} (a^{\dagger} m+a m^{\dagger})+g_{\rm mb} m^\dagger m  \left (b+ b^\dagger\right)  + K_{\rm{m}} m^{\dagger} m m^{\dagger} m   \\
&\! + K_{\rm cross}  m^\dagger m b^\dagger b +\sqrt{\kappa_{\rm{1}}}\varepsilon_{\rm{d}}\left(a e^{i \omega_{\rm{d}} t}+a^{\dagger} e^{-i \omega_{\rm{d}} t}\right)+\sqrt{\kappa_{\rm{2}}}\varepsilon_{\rm{p}}\left(a e^{i \omega_{\rm{p}} t}+a^{\dagger} e^{-i \omega_{\rm{p}} t}\right),
\end{aligned}
\end{equation}
where $\kappa_{\rm{1}}$ ($\kappa_{\rm{2}}$) is the cavity decay rate associated with the driving (probe) port, $\varepsilon_{\rm{d}}=\sqrt{\frac{P_{\rm{d}}}{\hbar\omega_{\rm{d}}}}$ and $\varepsilon_{\rm{p}}= \sqrt{\frac{P_{\rm{p}}}{\hbar\omega_{\rm{p}}}}$, with $P_{\rm{d}}$ ($P_{\rm{p}}$) and $\omega_{\rm{d}}$ ($\omega_{\rm{p}}$) being the power and frequency of the drive (probe) microwave field. Since in the experiment the probe field is of a much smaller power than that of the drive field and thus can be treated as a perturbation to the system, we therefore omit the probe term in the Hamiltonian, giving rise to the Hamiltonian (1) that is provided in the main text
\begin{equation}\label{sup13}
\begin{aligned}
H/\hbar \, =\, &\omega_{\rm{a}} a^{\dagger} a+\omega_{\rm{m}} m^{\dagger} m+\omega_{\rm{b}} b^{\dagger} b+g_{\rm ma} (a^{\dagger} m+a m^{\dagger})+g_{\rm mb} m^\dagger m  \left (b+ b^\dagger\right)    \\
&\! + K_{\rm{m}} m^{\dagger} m m^{\dagger} m + K_{\rm cross}  m^\dagger m b^\dagger b +\sqrt{\kappa_{\rm{1}}}\varepsilon_{\rm{d}}\left(a e^{i \omega_{\rm{d}} t}+a^{\dagger} e^{-i \omega_{\rm{d}} t}\right).
\end{aligned}
\end{equation}

\section{\uppercase\expandafter{\romannumeral4}.~Determination of the magnon excitation number}
\label{333}

From the Hamiltonian \eqref{sup13}, we can obtain the Heisenberg-Langevin equations by including the dissipation {and input noise} of each mode. 
In the frame rotating at the drive frequency, they are given by
\begin{equation}\label{sup14}
\begin{aligned}
&\frac{da}{dt}=- \left(i\Delta_{\rm{a}}+ \frac{\kappa_{\rm{a}}}{2} \right) a-ig_{\rm{ma}} m-i \sqrt{ \kappa_{\rm{1}}} \varepsilon_{\rm{d}}  \, {+ \sqrt{\kappa_{\rm{a}}}a_{\rm {in}} }, \\
&\frac{dm}{dt}=-\left[i \left(\Delta_{\rm{m}}+2 K_{\rm{m}}m^{\dagger}m +K_{\rm{m}}+ K_{\rm{cross}}b^{\dagger}b \right) +ig_{\rm{mb}}(b+b^{\dagger})+ \frac{\kappa_{\rm{m}}}{2}\right] m -ig_{\rm{ma}} a \, {+\sqrt{\kappa_{\rm{m}}}m_{\rm {in}} }, \\
& \frac{db}{dt}=-\left[i \left(\omega_{\rm{b}}+K_{\rm{cross}}m^{\dagger}m \right)+ \frac{\kappa_{\rm{b}}}{2} \right]b -ig_{\rm{mb}}m^{\dagger}m  \, {+\sqrt{\kappa_{\rm{b}}}b_{\rm {in}}},
\end{aligned}
\end{equation}
where $\Delta_{\rm{a}}=\omega_{\rm{a}}-\omega_{\rm{d}}$, and $\Delta_{\rm{m}}=\omega_{\rm{m}}-\omega_{\rm{d}}$, while $\kappa_{\rm{a}}$, $\kappa_{\rm{m}}$ and $\kappa_{\rm{b}}$ {($a_{\rm {in}}$, $m_{\rm {in}}$, and $b_{\rm {in}}$)} are the dissipation rates  {(input noises)} of the three modes.  {Since the cavity mode is strongly driven, this leads to a large amplitude $|\langle a \rangle| \gg 1$ in the steady state, and further due to the cavity-magnon coupling, the magnon mode also has a large amplitude $|\langle m \rangle| \gg 1$. This allows us to linearize the system dynamics around the classical average values by writing the mode operators as $a\equiv A+\delta a$, $m\equiv M+\delta m$, and $b\equiv B+\delta b$, and neglecting small second-order fluctuation terms~\cite{DVV}. Substituting these mode operators into Eq.~\eqref{sup14}, the equations are then separated into two sets of equations, respectively, for classical averages ($A$, $M$, $B$) and for quantum fluctuations ($\delta a$, $\delta m$, $\delta b$).}
The equations for the classical averages {\it in the steady state} are as follows:
\begin{equation}\label{sup15}
\begin{aligned}
&\left( \Delta_{\rm{a}}-i\frac{\kappa_{\rm{a}}}{2} \right) A+g_{\rm{ma}} M+ \sqrt{ \kappa_{\rm{1}}} \varepsilon_{\rm{d}}=0, \\
&\left[\Delta_{\rm{m}}+2 K_{\rm{m}}|M|^{2} +K_{\rm{m}}+ K_{\rm{cross}}|B|^{2} +g_{\rm{mb}}(B+B^{*})-i \frac{\kappa_{\rm{m}}}{2} \right] M +g_{\rm{ma}} A=0, \\
&\left(\omega_{\rm{b}}+K_{\rm{cross}}|M|^{2}-i \frac{\kappa_{\rm{b}}}{2} \right)B +g_{\rm{mb}}|M|^{2}=0.
\end{aligned}
\end{equation}

From the first equation of Eq.~\eqref{sup15}, we get
\begin{equation}\label{sup16}
\begin{aligned}
 A=-\eta_{\rm{a}}g_{\rm{ma}}\left( \Delta_{\rm{a}}+i \frac{\kappa_{\rm{a}}}{2} \right) M -\eta_{\rm{a}} \sqrt{ \kappa_{\rm{1}}} \left( \Delta_{\rm{a}}+i \frac{\kappa_{\rm{a}}}{2} \right) \varepsilon_{\rm{d}},
\end{aligned}
\end{equation}
where $\eta_{\rm{a}}= \frac{1} {\Delta_{\rm{a}}^2+(\frac{\kappa_{\rm{a}}}{2})^2}$. Substituting $A$ into the second equation of Eq.~\eqref{sup15}, we obtain
\begin{equation}\label{sup17}
\begin{aligned}
\left[\Delta_{\rm{m}} \! - \eta_{\rm{a}} g_{\rm{ma}}^2 \Delta_{\rm{a}} \,{+}\, 2 K_{\rm{m}}|M|^{2} {+}\,K_{\rm{m}} {+} \,K_{\rm{cross}}|B|^{2} {+} \, 2g_{\rm{mb}} \rm{Re}[B]   - i \left(\frac{\kappa_{\rm{m}}}{2}+\eta_{\rm{a}}g_{\rm{ma}}^2 \frac{\kappa_{\rm{a}}} {2} \right)\right] M + \eta_{\rm{a}} \sqrt{ \kappa_{\rm{1}}} g_{\rm{ma}} (\Delta_{\rm{a}}+i \frac{\kappa_{\rm{a}}}{2}) \varepsilon_{\rm{d}}=0.
\end{aligned}
\end{equation}
Multiplying Eq.~\eqref{sup17} with its complex conjugate, we obtain the equation for the magnon excitation number
\begin{equation}\label{sup18}
\left[\left(\Delta_{\rm{m}}-\eta_{\rm{a}}g_{\rm{ma}}^2 \Delta_{\rm{a}} +2 K_{\rm{m}}|M|^{2} +K_{\rm{m}}+ K_{\rm{cross}}|B|^{2} +2g_{\rm{mb}} \rm{Re}[B] \right)^2+ \left(\frac{\kappa_{\rm{m}}}{2}+\eta_{\rm{a}}g_{\rm{ma}}^2 \frac{\kappa_{\rm{a}}} {2} \right)^2 \right]|M|^{2} =\eta_{\rm{a}} \kappa_{\rm{1}} g_{\rm{ma}}^2 \varepsilon_{\rm{d}}^2.
\end{equation}
Similarly, we get the equation for the phonon excitation number
\begin{equation}\label{sup19}
\left[\left(\omega_{\rm{b}}+K_{\rm{cross}}|M|^{2}\right)^{2}+\left( \frac{\kappa_{\rm{b}}}{2} \right)^{2}\right]|B|^{2} =g_{\rm{mb}}^{2}|M|^{4}.
\end{equation}

In our experiment, the drive power is scanned from 4.7 dBm to 23.7 dBm, which gives the magnon excitation number $|M|^{2} \in [10^{12},10^{15}]$, and the phonon excitation number $|B|^{2} \in [10^{6},10^{10}]$. Thus, the phonon excitation number is much smaller than that of the magnon. Using our parameters $g_{\rm{mb}}/2\pi=1.22~\rm{mHz}$, $K_{\rm{m}}/2\pi=-6.5~\rm{nHz}$, and $K_{\rm{cross}}/2\pi=-5.4~\rm{pHz}$, the magnon frequency shift $\delta \omega_{\rm{m}}= 2 K_{\rm{m}}|M|^{2}+K_{\rm{m}}+ K_{\rm{cross}}|B|^{2} +2 g_{\rm{mb}}{\rm Re}[B] \approx 2 K_{\rm{m}}|M|^{2}$.  Therefore, Eq.~\eqref{sup18} is reduced to
\begin{equation}\label{sup20}
\left[\left(\Delta_{\rm{m}}-\eta_{\rm{a}}g_{\rm{ma}}^2 \Delta_{\rm{a}} +2 K_{\rm{m}}|M|^{2} \right)^2+ \left(\frac{\kappa_{\rm{m}}}{2}+\eta_{\rm{a}}g_{\rm{ma}}^2 \frac{\kappa_{\rm{a}}} {2} \right)^2 \right]|M|^{2} =\eta_{\rm{a}} \kappa_{\rm{1}} g_{\rm{ma}}^2 \varepsilon_{\rm{d}}^2.
\end{equation}
This is a cubic equation of the magnon excitation number $|M|^{2}$, and given as Eq. (4) in the main text. Under certain conditions, all the three solutions of $|M|^{2}$ are real, among which there are two stable solutions. The stable solutions can be measured in the experiment, and it shows a hysteresis loop by varying the drive power.

\section{\uppercase\expandafter{\romannumeral5}.~Effective susceptibility of the mechanical mode}
\label{444}

The magnon-phonon {radiation-pressure-like} coupling gives rise to the magnomechanical backaction on the mechanical mode, which is manifested as the mechanical frequency shift (i.e., the magnonic spring effect) and the increased (reduced) linewidth associated with the cooling (amplification) of the mechanical mode. The frequency shift and the linewidth variation can be evaluated from the effective susceptibility of the mechanical mode. In what follows, we show in detail how the effective mechanical susceptibility is derived.

{The linearization of the Langevin equations~\eqref{sup14} yields a set of linearized quantum Langevin equations for the quantum fluctuations $(\delta m, \delta a, \delta x, \delta p)$,} where $\delta x=(\delta b+\delta b^\dagger)/\sqrt{2}$ and $\delta p=i (\delta b^\dagger - \delta b)/\sqrt{2}$ denote the fluctuations of two mechanical quadratures (position and momentum).  {By taking the Fourier transform, we obtain the following equations in the frequency domain:}
\begin{equation}\label{sup21}
\begin{aligned}
&-i\omega \delta m=-\left ( i\tilde{\Delta}_{\rm{m}}+\frac{\kappa_{\rm{m}}}{2} \right ) \delta m -ig_{\rm{ma}}\delta a -i\sqrt{2} G_{{\rm{mb}}}\delta x +\sqrt{\kappa_{\rm{m}}}m_{\rm {in}}, \\
&-i\omega \delta m^{\dagger}=-\left (-i\tilde{\Delta}_{\rm{m}}+\frac{\kappa_{\rm{m}}}{2} \right ) \delta m^{\dagger} +ig_{\rm{ma}}\delta a^{\dagger} +i\sqrt{2} G_{{\rm{mb}}}^{*}\delta x + \sqrt{\kappa_{\rm{m}}}m_{\rm{in}}^{\dagger}, \\
&-i\omega \delta a=-\left ( i\Delta_{\rm{a}}+\frac{\kappa_{\rm{a}}}{2} \right )\delta a -ig_{{\rm{ma}}}\delta m +\sqrt{\kappa_{\rm{a}}}a_{\rm{in}},  \\
&-i\omega \delta a^{\dagger}=-\left (-i\Delta_{\rm{a}}+\frac{\kappa_{\rm{a}}}{2} \right )\delta a^{\dagger} +ig_{{\rm{ma}}}\delta m^{\dagger} +\sqrt{\kappa_{\rm{a}}}a_{\rm{in}}^{\dagger},  \\
&-i\omega\delta x= \tilde{\omega}_{\rm{b}}\delta p,\\
&-i\omega\delta p= -\tilde{\omega}_{\rm{b}} \delta x - \kappa_{\rm{b}} \delta p - \sqrt{2} \left( G_{{\rm{mb}}}^{*} \delta m +G_{{\rm{mb}}}\delta m^{\dagger} \right) +\xi,
\end{aligned}
\end{equation}
where $\tilde{\Delta}_{\rm{m}} \simeq \Delta_{\rm{m}}+2 K_{\rm{m}}|M|^{2}$ includes the magnon frequency shift {\it dominantly} caused by the magnon self-Kerr effect, and $\tilde{\omega}_{\rm{b}}=\omega_{\rm{b}}+ K_{\rm{cross}}|M|^{2}$ includes the mechanical frequency shift due to the magnon-phonon cross-Kerr effect. $G_{{\rm{mb}}}=g_{{\rm{mb}}}M$ is the effective {radiation-pressure-like} coupling strength. {Note that we adopt an equivalent model of dealing with the mechanical damping and input noise, where the damping rate $\kappa_b$ and the Hermitian Brownian noise operator $\xi$ are added only in the momentum equation~\cite{DVV}}. Solving separately the two equations for each mode, we obtain the following equations:
\begin{equation}\label{sup22}
\begin{aligned}
&\delta m=\chi_{\rm{m}}(\omega) \left(-ig_{\rm{ma}}\delta a-i\sqrt{2} G_{{\rm{mb}}}\delta x+\sqrt{\kappa_{\rm{m}}}m_{\rm {in}}\right),\\
&\delta m^{\dagger}=\chi_{\rm{m}}^{*}(-\omega)\left(ig_{\rm{ma}}\delta a^{\dagger}+i\sqrt{2} G_{{\rm{mb}}}^{*}\delta x +\sqrt{\kappa_{\rm{m}}}m_{\rm {in}}^{\dagger}\right),\\
&\delta a=\chi_{\rm{a}}(\omega)\left(-ig_{\rm{ma}}\delta m +\sqrt{\kappa_{\rm{a}}}a_{\rm{in}}\right), \\
&\delta a^{\dagger}=\chi_{\rm{a}}^{*}(-\omega)\left(ig_{\rm{ma}}\delta m^{\dagger} +\sqrt{\kappa_{\rm{a}}} a_{\rm{in}}^{\dagger}\right), \\
&\delta x =\chi_{\rm{b}}(\omega) \left(- \sqrt{2} G_{{\rm{mb}}}^{*} \delta m - \sqrt{2}G_{{\rm{mb}}} \delta m^{\dagger}  +\xi \right),\\
&\delta p=\frac{-i\omega}{\tilde{\omega}_{\rm{b}}}\delta x,
\end{aligned}
\end{equation}
where we define the natural susceptibilities of the magnon, cavity, and mechanical modes as
\begin{equation}\label{sup23}
\begin{aligned}
&\chi_{\rm{m}}(\omega)=\frac{1}{i\left(\tilde{\Delta}_{\rm{m}}-\omega \right)+\frac{\kappa_{\rm{m}}}{2}}, \,\,\, \chi_{\rm{m}}^{*}(-\omega)=\frac{1}{-i\left(\tilde{\Delta}_{\rm{m}}+\omega \right)+ \frac{\kappa_{\rm{m}}}{2}},\\
&\chi_{\rm{a}}(\omega)=\frac{1}{i\left(\Delta_{\rm{a}}-\omega \right)+\frac{\kappa_{\rm{a}}}{2}}, \,\,\,\,\,\,
\chi_{\rm{a}}^{*}(-\omega) =\frac{1}{-i\left(\Delta_{\rm{a}}+\omega \right)+ \frac{\kappa_{\rm{a}}}{2}},\\
&\chi_{\rm{b}}(\omega)=\frac{\tilde{\omega}_{\rm{b}}}{\tilde{\omega}_{\rm{b}}^{2}-\omega^{2}-i\kappa_{\rm{b}}\omega }.
\end{aligned}
\end{equation}

Solving the first four equations in Eq.~\eqref{sup22} for $\delta m$ and $\delta m^{\dagger}$, and inserting their solutions into the equation of $\delta x$, we obtain
\begin{equation}\label{sup24}
 \delta x= \chi_{\rm{b,eff}}(\omega) \!\!  \left( \! \xi {+} \frac{i\sqrt{2}G_{\rm{mb}}^{*}\chi_{\rm{m}}(\omega) \! \left( i\sqrt{\kappa_{\rm{m}}} m_{\rm{in}} \!+ \!\! \sqrt{\kappa_{\rm{a}}} g_{\rm{ma}}\chi_{\rm{a}}(\omega) a_{\rm {in}}\right) } {1+ g^2_{\rm{ma}} \chi_{\rm{a}}(\omega)\chi_{\rm{m}}(\omega) } {-} \frac{ i\sqrt{2}G_{\rm{mb}}\chi_{\rm{m}}^{*}(-\omega) \! \left( \! -i\sqrt{\kappa_{\rm{m}}} m_{\rm{in}}^{\dagger} \!+\!\! \sqrt{\kappa_{\rm{a}}} g_{\rm{ma}}\chi_{\rm{a}}^{*}(-\omega) a_{\rm{in}}^{\dagger}\right) } {1+ g^2_{\rm{ma}} \chi_{\rm{a}}^{*}(-\omega)\chi_{\rm{m}}^{*}(-\omega) } \!  \right),
\end{equation}
where $\chi_{\rm{b,eff}}$ is the effective mechanical susceptibility, defined as
\begin{equation}\label{sup25}
\begin{aligned}
\chi_{\rm{b,eff}}(\omega)
=\left( \chi_{\rm{b}}^{-1}(\omega)-2i|G_{\rm{mb}}|^{2}\left( \chi_{\rm{ma}}(\omega)- \chi_{\rm{ma}}^{*}(-\omega) \right) \right)^{-1},
\end{aligned}
\end{equation}
with $\chi_{\rm{ma}}(\omega)=\frac{1}{\chi_{\rm{m}}^{-1}(\omega)+g_{\rm{ma}}^2 \chi_{\rm{a}}(\omega)}$. The change of the mechanical frequency and linewidth can be extracted from the effective susceptibility: the real part of $\chi_{\rm{b,eff}}^{-1}(\omega)-\chi_{\rm{b}}^{-1}(\omega)$ corresponds to the mechanical frequency shift
\begin{equation}\label{sup26}
\delta\omega_{\rm{b}}= -{\rm{Re}}\left[ 2i|G_{\rm{mb}}|^{2}\left( \chi_{\rm{ma}}(\omega)- \chi_{\rm{ma}}^{*}(-\omega)\right) \right]+ K_{\rm{cross}} |M|^{2},
\end{equation}
where we write together the frequency shift due to the cross-Kerr effect, and the imaginary part of $\chi_{\rm{b}}^{-1}(\omega) - \chi_{\rm{b,eff}}^{-1}(\omega)$ yields the variation of the mechanical linewidth
\begin{equation}\label{sup27}
\delta\Gamma_{\rm{b}}={\rm{Im}}\left[ 2i|G_{\rm{mb}}|^{2}\left( \chi_{\rm{ma}}(\omega)- \chi_{\rm{ma}}^{*}(-\omega)\right) \right].
\end{equation}

\section{\uppercase\expandafter{\romannumeral6}.~Reflection spectrum of the probe field}

Here we show how to derive the reflection spectrum of the probe field under the strong drive field. The Hamiltonian including the probe field is given in Eq.~\eqref{sup12}.  The reflection spectrum can be conveniently solved by including the strong pump effects into the linearized Langevin equations. {Following the linearization approach used in Sec. IV and Sec. V, the Hamiltonian~\eqref{sup12} leads to the following Langevin equations for the classical averages in the frequency domain:}
\begin{equation}\label{sup28}
\begin{aligned}
&-i\omega M=-\left ( i\tilde{\Delta}_{\rm{m}}+\frac{\kappa_{\rm{m}}}{2} \right ) M -ig_{\rm{ma}} A -i\sqrt{2} G_{{\rm{mb}}} X,\\
&-i\omega A=-\left ( i\Delta_{\rm{a}}+\frac{\kappa_{\rm{a}}}{2} \right ) A -ig_{{\rm{ma}}} M -i\sqrt{\kappa_{\rm{2}}}\varepsilon_{\rm{p}} \delta(\omega_{\rm{p}}-\omega_{\rm{d}}-\omega ),  \\
&-i\omega X= \tilde{\omega}_{\rm{b}} P,\\
&-i\omega P= -\tilde{\omega}_{\rm{b}} X- \kappa_{\rm{b}} P - \sqrt{2} G_{{\rm{mb}}}^{*} M,
\end{aligned}
\end{equation}
{where $X=(B+B^*)/\sqrt{2}$ and $P=i (B^* - B)/\sqrt{2}$ denote the classical averages of the mechanical position and momentum. Note that, same as Eqs.~\eqref{sup14} and \eqref{sup21}, the above equations are provided in the frame rotating at the drive frequency $\omega_{\rm{d}}$. }

Solving the above equations, we obtain
\begin{equation}\label{sup29}
\begin{aligned}
A(\omega) = \frac{i\sqrt{\kappa_{\rm{2}}} \left(1-2i|G_{\rm{mb}}|^{2}\chi_{\rm{b}}(\omega)\chi_{\rm{m}}(\omega) \right)} {-g_{\rm{mb}}^{2} (\omega)\chi_{\rm{m}}(\omega) - \chi^{-1}_{\rm{a}}(\omega) \big( 1-2i|G_{\rm{mb}}|^{2}\chi_{\rm{b}}(\omega)\chi_{\rm{m}}(\omega) \big) } \varepsilon_{\rm{p}}.
\end{aligned}
\end{equation}

Using the input-output theory, $A_{\rm{out}}=\varepsilon_{\rm{p}}+i\sqrt{\kappa_{\rm{2}}}A $, we therefore achieve the reflection spectrum of the probe field
\begin{equation}\label{sup30}
\begin{aligned}
r(\omega) \equiv \frac{A_{\rm{out}}}{\varepsilon_{\rm{p}}} = 1- \frac{\kappa_{\rm{2}} \left(1-2i|G_{\rm{mb}}|^{2}\chi_{\rm{b}}(\omega)\chi_{\rm{m}}(\omega) \right)} {-g_{\rm{mb}}^{2} (\omega)\chi_{\rm{m}}(\omega) - \chi^{-1}_{\rm{a}}(\omega) \big( 1-2i|G_{\rm{mb}}|^{2}\chi_{\rm{b}}(\omega)\chi_{\rm{m}}(\omega) \big) }.
\end{aligned}
\end{equation}

\end{document}